\documentstyle[aps,prl,floats,graphicx]{revtex}
\begin{document}
\draft
\twocolumn[\hsize\textwidth\columnwidth\hsize\csname @twocolumnfalse\endcsname
\title{Critical Velocity of Vortex Nucleation in Rotating Superfluid
$^3$He-A}    
\author{V.M.H. Ruutu$^{(1)}$, J. Kopu$^{(1)}$, M. Krusius$^{(1)}$, 
\"U. Parts$^{(1)}$, B.
Pla\c{c}ais$^{(2)}$,  E.V. Thuneberg$^{(1)}$, and W. Xu$^{(1)}$}
\address{$^{(1)}$Low Temperature Laboratory, Helsinki University of Technology,
FIN-02015 HUT, Finland\\ $^{(2)}$Ecole Normale Sup\'erieure, CNRS URA
1437, 24 rue Lhomond, F-75231 Paris Cedex 05, France} 
\date{\today}
\maketitle
\begin{abstract}
We have measured the critical velocity $v_{\rm c}$ at which $^3$He-A in a
rotating cylinder becomes unstable against the formation of quantized vortex
lines with continuous (singularity-free) core structure. We find that $v_{\rm
c}$ is distributed between a maximum and minimum limit, which we ascribe to a
dependence on the texture of the orbital angular momentum $\hat{\bf l}({\bf r})$
in the cylinder. Slow cool down through
$T_{\rm c}$ in rotation yields $\hat{\bf l}({\bf r})$ textures for which the
measured
$v_{\rm c}$'s are in good agreement with the calculated instability of the
expected
$\hat{\bf l}$ texture.
\end{abstract} 
   \bigskip 
] 
 
A first order transition from one phase to another is associated with
hysteresis because of the difficulty of nucleating the new phase. Two effects
generally reduce the hysteresis. Firstly, thermal or quantum fluctuations cause
the new phase to appear before the energy barrier separating the two energy
minima vanishes. Secondly, surfaces, impurities, or other external agents reduce
the energy barrier from its intrinsic value. Both phenomena are of crucial
importance for the long standing problem of critical velocities and vortex
nucleation in superfluids \cite{Donnelly}, but occur also in
more usual phenomena like formation of water droplets or gas bubbles
\cite{bubbles}. The purpose of the present work is to study an exceptional case
of vortex nucleation where neither fluctuations nor external surfaces should
play a role: superfluid $^3$He-A.

In usual superfluids and superconductors the phase slip takes place by creation
and motion of zeros in the order parameter \cite{Anderson}. The A phase of $^3$He
is exceptional because the phase slip arises from the motion of the local angular
momentum axis $\hat{\bf l}({\bf r})$. The characteristic length scale of the
$\hat{\bf l}({\bf r})$ texture is macroscopic $\sim 10\ \mu$m. Therefore all
thermal and quantum fluctuations are negligible. Moreover, a rigid boundary
condition fixes $\hat{\bf l}$ perpendicular to the wall of the experimental
container. Thus the processes responsible for the critical velocity take place
further than $ 10\ \mu$m from the wall, beyond the reach of surface roughness.
Instead, the critical velocity $v_{\rm c}$ for the phase slip depends on the
initial $\hat{\bf l}$ texture. We measure the critical velocity in a rotating
cylinder, and find that it may vary within a factor of 6. However, by cooling
slowly through the superfluid transition temperature $T_{\rm c}$ in rotation,
the equilibrium texture is created and the measured $v_{\rm c}$ is 
in agreement with theoretical calculations. 

{\it Anisotropic superflow.}---In ordinary superconductors and superfluids, the
order parameter has a phase factor $\exp[i\phi({\bf r})]$, and the superfluid
velocity is defined as the gradient of the phase, ${\bf v}_{\rm
s}\propto\bbox{\nabla}\phi$. In $^3$He-A there is an additional phase factor
$\exp[i\phi_l(\hat{\bf p})]$, which depends on the azimuthal angle $\phi_l$ of
the quasiparticle momentum ${\bf p}$ with respect to the angular momentum axis
$\hat{\bf l}$. Instead of resolving the two phases separately, one may only
define the total phase factor, which can be expressed as $(\hat{\bf m}+i\hat{\bf
n})\cdot\hat{\bf p}$. Here $\hat{\bf l}$, $\hat{\bf m}$, and $\hat{\bf n}$ form
an orthonormal triad, which generally depends on the location ${\bf r}$. The
superfluid velocity is defined as ${\bf v}_{\rm s}={\hbar\over 2m}\sum_k\hat
m_k\bbox{\nabla}\hat n_k$, where the prefactor ${\hbar\over 2m}$ equals Planck's
constant divided by twice the mass of a $^3$He atom. This leads to several
unusual features. For example, let us take an initially uniform
$\hat{\bf l}\equiv\hat{\bf z}$, and then tilt the triads so that $\hat{\bf
l}(z)$ forms a helix with an opening angle $\beta$ and a wave vector $q$. This
leads to a change in $v_{{\rm s},z}$ by ${\hbar\over 2m}(1-\cos\beta)q$ without
a change in the externally applied phase difference \cite{BHM}. Thus $^3$He-A
can respond to flow by forming an $\hat{\bf l}$ texture.

The energetics of the current carrying states is contained in the energy
functional \cite{VolWol}
\begin{eqnarray}
f&=&\textstyle{1\over 2}
\rho_\perp{\bf w}^2+\textstyle{1\over 2}(\rho_\parallel-\rho_\perp)(\hat{\bf
l}\cdot{\bf w})^2
\nonumber \\&\hbox{}-&C{\bf w}\cdot\nabla\times\hat{\bf l} +C_0(\hat{\bf
l}\cdot{\bf w}) (\hat{\bf l}\cdot\nabla\times\hat{\bf l})\nonumber \\&\hbox{}+&
\textstyle{1\over 2}K_{\rm s}(\nabla\cdot\hat{\bf l})^2 +
\textstyle{1\over 2}K_{\rm t}(\hat{\bf
l}\cdot\nabla\times\hat{\bf l})^2+\textstyle{1\over 2}K_{\rm b}\vert\hat{\bf
l}\times(\nabla\times\hat{\bf l})\vert^2 \nonumber \\ &\hbox{}+&
\textstyle{1\over 2}K_5 \vert(\hat{\bf
l}\cdot\nabla)\hat{\bf d}\vert^2+ \textstyle{1\over 2}K_6[(\hat{\bf
l}\times\nabla)_i\hat{\bf d}_j)]^2\nonumber \\& \hbox{}
-&\textstyle{1\over 2}g_{\rm
d}(\hat{\bf d}\cdot\hat{\bf l})^2
+\textstyle{1\over 2}g_{\rm h}(\hat{\bf d}\cdot{\bf
H})^2\ . \label{e.f}
\end{eqnarray}
Here the first two terms describe the anisotropic kinetic energy arising from
the counterflow ${\bf w}={\bf v}_{\rm n}-{\bf v}_{\rm s}$. The terms with
coefficients $C$ and $C_0$ contribute to the coupling between ${\bf w}$ and
inhomogeneous $\hat{\bf l}({\bf r})$. The five terms with $K_i$ coefficients are
gradient energies for $\hat{\bf l}({\bf r})$ and the spin anisotropy vector
$\hat{\bf d}({\bf r})$. The last two terms arise from the magnetic dipole-dipole
interaction, and from the external magnetic field ${\bf H}$. In the absence of
the dipole coupling, all uniform current carrying states would be unstable
\cite{BHM}. Thus the dipole interaction determines the scale of the critical
velocity $v_{\rm d}=\sqrt{g_{\rm d}/\rho_\parallel}\sim 1$ mm/s and the length
scale $\xi_{\rm d}={\hbar\over 2m}\sqrt{\rho_\parallel/g_{\rm d}}\sim 10\ \mu$m.
Except for a tiny region near $T_{\rm c}$ (Fig.\
2), $v_{\rm d}$ is much smaller than is
needed to nucleate usual ``singular vortices'', such as one encounters in other
superfluids \cite{Donnelly,SingleVortexLett}.  

Let us drive the current by applying a normal fluid velocity $v_{\rm n}$. (This
is equivalent to applying a phase difference $\Delta\phi={2m\over\hbar}v_{\rm
n}L$ between two points a distance $L$ apart.) The flow properties depend on the
magnitude and orientation of the magnetic field \cite{VLM,LVM,FW,F81}. In
general, a uniform state of $\hat{\bf l}$ is stable for velocities smaller than
a first critical velocity
$v_{\rm c1}$. At larger velocities there often is a stable helical texture. The
opening angle $\beta$ of the helix grows continuously from zero with increasing
$v_{\rm n}$ until a second critical velocity $v_{\rm c2}$. There the helix
becomes unstable, and the resulting state depends on how the flow is applied. A
continuously sustained dissipative $\hat{\bf l}$ texture is found when a
constant current is driven in a channel. This case has been studied by several
groups, most extensively by Bozler and collaborators
\cite{PaalanenOsheroff,Bozler,Saundry}. In our rotating cylinder, a stationary
state is restored after one or more vortex lines are formed in a phase slip.  

{\it Experiment.}---Our sample container is a cylinder which is rotated around
its axis with angular velocity $\Omega$. The radius of the cylinder $R\sim 2$ mm
is large compared to $\xi_{\rm d}$, which means that the normal velocity ${\bf
v}_{\rm n}=\bbox{\Omega}\times{\bf r}$ is rectilinear to a good approximation
near the cylindrical walls. Therefore, one should see transitions at
$\Omega_{\rm c1}=v_{\rm c1}/R$ and  $\Omega_{\rm c2}=v_{\rm c2}/R$ corresponding
to the critical values of one-dimensional flow. At
$\Omega_{\rm c2}$ dissipation sets in only temporarily when a vortex line is 
created. It is driven by the Magnus force into a vortex bundle in the center of
the cylinder. As a consequence, the effective driving velocity $v=
[\Omega-\Omega_{\rm v}(N)]R$ is reduced to a subcritical value. Here $N$ is the
number of vortices in the bundle, $\Omega_{\rm v} = \kappa N/2\pi R^2$, and
$\kappa$ the circulation of one vortex line. Compared to channel flow, our
experiment has high resolution in the measurement of $v_{\rm c2}$, since the
vortices can be counted with a precision of $\pm 5$ lines from the
continuous-wave NMR spectrum.  Our sensitivity to a helical texture is poor and
$v_{\rm c1}$ has not been observed.
   
Three different sample cylinders have been used, which were fabricated from
epoxy or fused quartz  with  radii $R = 2$ -- 2.5 mm, heights $L = 6$ -- 7 mm,
and different surface roughness \cite{SingleVortexLett}. No systematic
dependence of $v_{\rm c}$ on the container was found. A small orifice in the
center of the bottom plate of the cylinder provides thermal contact via a liquid
$^3$He column to the refrigerator.  The NMR field ${\bf H}$, which is large
compared to the dipole field $H_{\rm d}=\sqrt{g_{\rm d}/g_{\rm h}}\approx 3$ mT,
is either axial ($\parallel{\bf\Omega}$) or transverse  ($\perp{\bf\Omega}$). The
NMR absorption spectrum has two peaks. The frequency shift of the main peak from
the Larmor value is used for thermometry. The satellite peak arises from vortex
lines \cite{CUV} and its intensity is proportional to the number of lines
$N$.  

\begin{figure}[tbp]
\begin{center}\leavevmode
\includegraphics[width=0.73\linewidth]{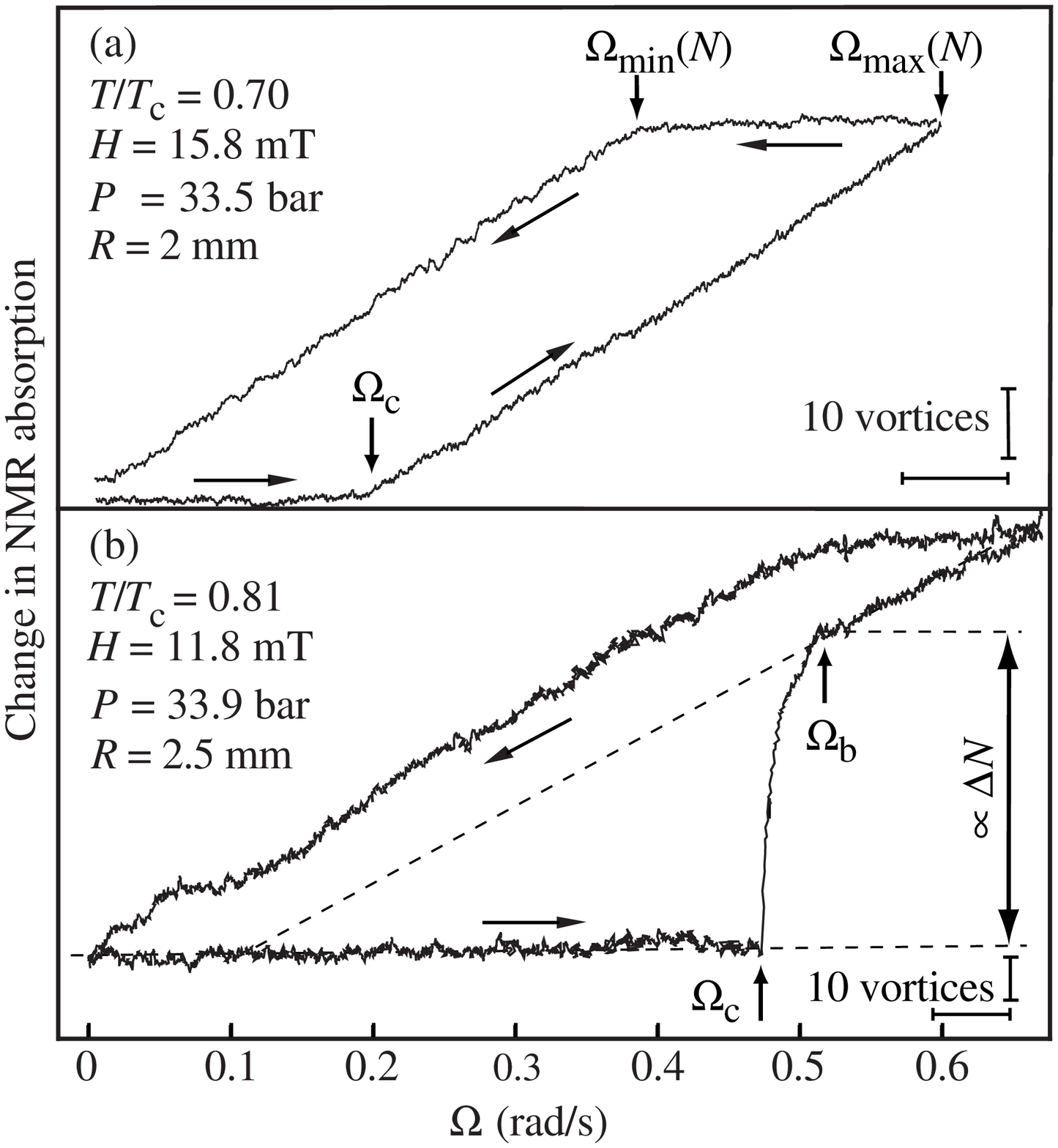}
\bigskip
\caption[fig:AccDec]{
NMR absorption at the vortex satellite peak during a slow
acceleration -- deceleration cycle: (a) regular vortex formation at
approximately constant $v_c$ and (b) burst-like vortex formation followed by the
regular process. The dashed line shows the extrapolation to $N\rightarrow 0$ in
the final state where the regular process has set in with a reduced $v_{\rm
c}$. 
}\label{fig:AccDec}\end{center}\end{figure}

{\it Results.}---Fig.\ \ref{fig:AccDec} shows two measurements of the amplitude
of the satellite  peak as a function of $\Omega$. The acceleration
is started from rest ($\Omega=N=0$) at a slow rate  ($d\Omega/dt = 10^{-3}$ --
$10^{-5}$ rad/s$^2$), and the satellite intensity remains zero until a critical
velocity $v_{\rm c} = \Omega_{\rm c} R$. In the top frame the amplitude starts
to increase linearly when $\Omega_{\rm c}$ is exceeded. This means that vortices
are nucleated regularly at a $v_{\rm c}$ that is approximately independent of
the number of vortices $N$ in the center of the cylinder.

The acceleration in Fig.\ 1a is stopped at a velocity $\Omega_{\rm max}$. In
order to determine $N$ and $v_{\rm c}$ in this state, the rotation is
decelerated until at $\Omega_{\rm min}(N)$ vortices start to annihilate
\cite{annihilation}. The regular parallelogram-like shape of the acceleration --
deceleration loop in  Fig.\ \ref{fig:AccDec}a shows that $v_{\rm c}$ is
approximately constant.

\begin{figure}[tbp]
\begin{center}\leavevmode
\includegraphics[width=0.8\linewidth]{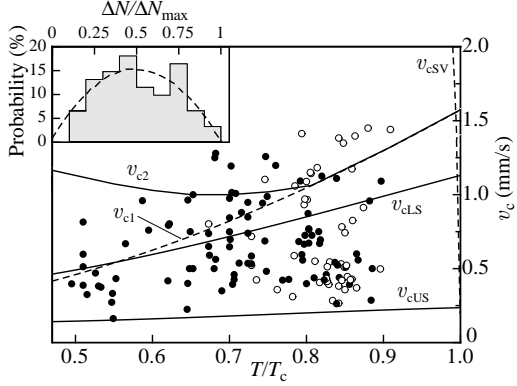}
\bigskip
\caption[f.all]{ 
 Critical velocities as a function of the temperature $T$ in axial field ($H =
9.9$ -- 15.8 mT, pressures 29.3 -- 34.2 bar). The 
experimental results, which represent a variety of sample histories, are
classified as regular ($\bullet$) and burst-like ($\circ$) vortex formation. The
solid lines denote instabilities towards vortex formation in different
one-dimensional textures:
$v_{\rm c2}$  of uniform or helical texture, $v_{\rm cLS}$ of a locked soliton
texture, and 
$v_{\rm cUS}$ of an unlocked soliton. The dashed lines show the
phase boundary between uniform and helical textures $v_{\rm c1}$, and the
instability to singular vortices $v_{\rm cSV}$ (estimated from Ref.\
\cite{AphDiag}).  {\it Inset:} Distribution of the number of vortices $\Delta N$
in a burst, normalized to the maximum possible increase
$\Delta N_{\rm max} = (2\pi R^2 / \kappa)(\Omega_{\rm b} - 0.16
\protect\sqrt{\Omega_{\rm b}}) - N_{\rm i}$, where $N_{\rm i}$ is the initial
vortex number and $\Omega_{\rm b}$ is defined in Fig.\ \protect\ref{fig:AccDec}.
}\label{f.all}\end{center}\end{figure}

On repeating the measurement, we generally find a considerable spread in $v_{\rm
c}$. The distribution is
evident in Fig.\ \ref{f.all}, which shows $v_{\rm c}$ measured for
different histories of sample preparation. Because this variation is much
larger than seen in one measuring run (Fig.\ 1a), it has to arise from different
metastable states of the system.  The only source of metastability in our system
are different superfluid states, i.e. different textures of $\hat{\bf l}$,
$\hat{\bf m}$,
$\hat{\bf n}$, and $\hat{\bf d}$.  Additional evidence for the textural origin
of the spread is discussed below.

In Fig.\ \ref{fig:AccDec}b the response to acceleration is a sudden jump in the
signal, which corresponds to a burst of $\Delta N \approx 90$ vortex lines. The
critical velocity  $v_{\rm c}$ after the jump is changed and is generally
smaller than before. The subsequent  deceleration to the
annihilation threshold in Fig.\ 1b shows that $v_{\rm c}$ is reduced from 1.2
mm/s to 0.26 mm/s. Obviously, this behavior is caused by a transition from one
texture to another. The vortex bursts ($\circ$ in Fig.\ 2) take place only at
high temperatures $T\gtrsim 0.7 \, T_{\rm c}$. This is consistent with the fact
that the energy barriers between different textures are smaller at higher
temperatures. 

Curve $v_{\rm c2}$ in Fig.\ \ref{f.all} is a theoretical result for the
vortex instability of an initially uniform texture. It roughly agrees with the
largest measured $v_{\rm c}$'s. Some measured values are larger than the
theoretical upper limit, which may arise from inaccurate parameter values in
the calculation (see below). In order to justify theoretically the
texture dependence of $v_{\rm c}$, we have calculated two simple
cases of initially inhomogeneous texture. A locked soliton (LS) is a planar
object where both $\hat{\bf d}$ and $\hat{\bf l}$ turn from parallel to ${\bf
v}$ on one side to antiparallel to
${\bf v}$ on the other side of the wall, while they are locked to each other
($\hat{\bf d}\equiv\hat{\bf l}$). Such a texture becomes unstable against vortex
formation at the velocity $v_{\rm cLS}$. Another type of domain wall is an
unlocked soliton (US), where  $\hat{\bf d}=\hat{\bf l}$ on one side and
$\hat{\bf d}=-\hat{\bf l}$ on the other. When its plane is perpendicular to
${\bf v}$, vortices are created at the velocity $v_{\rm cUS}$ \cite{SplaySol}.
These simple cases span the spread of the measured $v_{\rm c}$ in Fig.\ 2. This
makes it plausible that the three-dimensional texture and its defects in the
cylinder are responsible for all of the variation in $v_{\rm c}$.
 
\begin{figure}[btp]
\begin{center}\leavevmode
\includegraphics[width=0.73\linewidth]{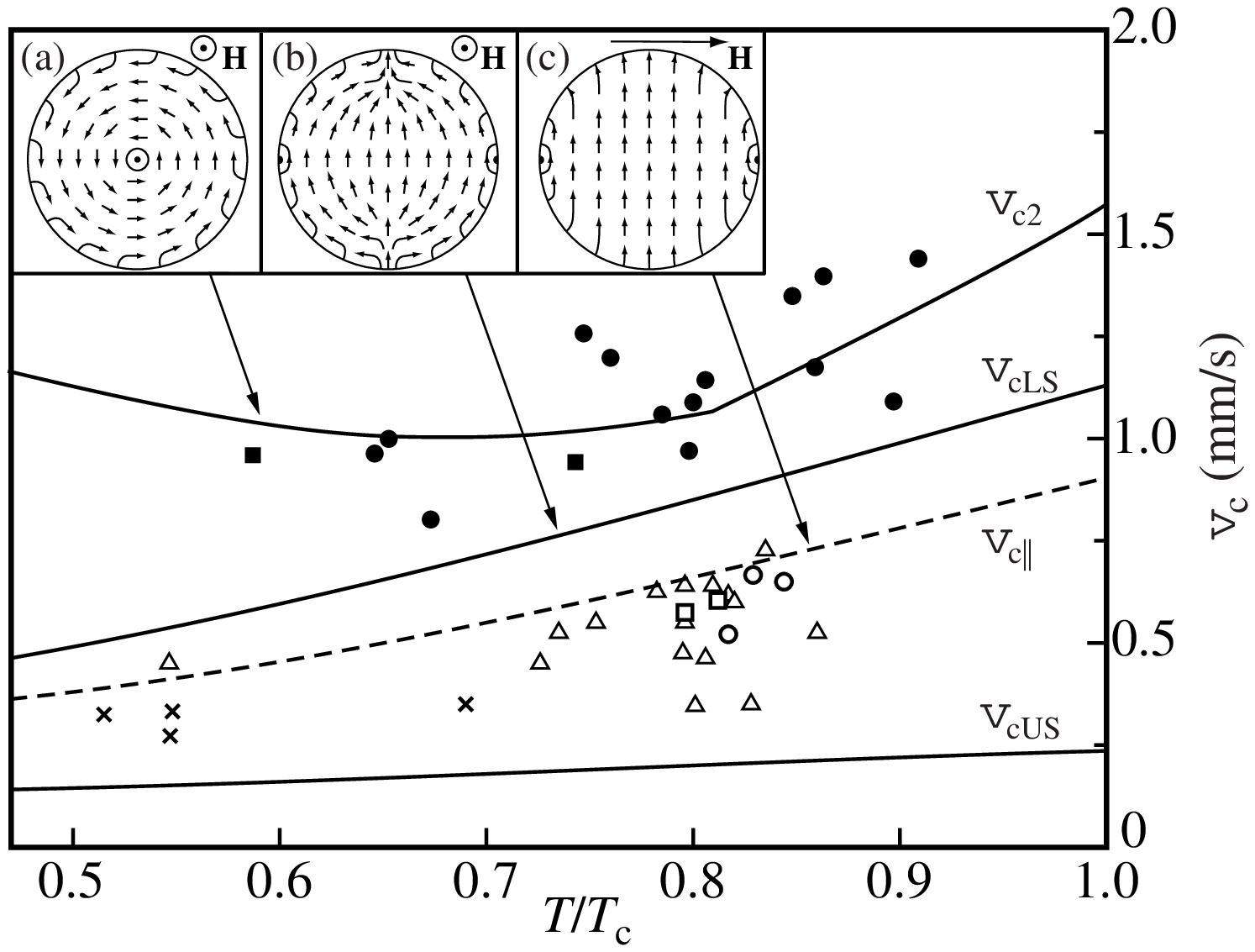}
\bigskip
\caption[f.selection]{ 
Critical velocities in selected cases. (i) We cool slowly through
$T_{\rm c}$ under rotation. In axial field this gives results ($\bullet$, \kern
.2ex\vrule height .95ex width .8ex depth -.15ex \kern .2ex) near the instability
limit $v_{\rm c2}$ of the circular texture (inset a). In transverse field
the corresponding results ($\circ$,
\kern .2ex\vrule height .98ex width .1ex depth -.18ex 
\vrule height .28ex width .7ex depth -.18ex 
\kern -.8ex \vrule height .98ex width .7ex depth -.88ex
\vrule height .98ex width .1ex depth -.18ex\kern .2ex)
are near the instability limit $v_{{\rm c}\parallel}$ \protect\cite{F81} of the 
expected texture (inset c). There is no crucial difference if the magnetic
field $H$ is on during cooling 
(\kern .2ex\vrule height .95ex width .8ex depth -.15ex \kern .2ex, 
\kern .2ex\vrule height .98ex width .1ex depth -.18ex 
\vrule height .28ex width .7ex depth -.18ex 
\kern -.8ex \vrule height .98ex width .7ex depth -.88ex
\vrule height .98ex width .1ex depth -.18ex\kern .2ex)  or it is switched on
later  ($\bullet$, $\circ$). The scatter in the data (which exceeds that of
$v_{\rm c}$ in Fig.\ 1a) indicates that the control over the texture is
limited, possibly due to the end plates of the cylinder.   (ii) In transverse
field
$v_{\rm c}$ seems to be less sensitive to the texture,
as shown by data ($\triangle$) with varied prehistories. (iii) Often a
transverse soliton plane ($\perp{\bf\Omega}$) is formed when cooling rapidly
through
$T_{\rm c}$ at
$\Omega = 0$. In these cases a low
$v_{\rm c}$ is found ($\times$, in axial field). 
}\label{f.selection}\end{center}\end{figure}

The data in Fig.\ \ref{f.all} has been collected under a variety of prehistories
of sample preparation. We now demonstrate that more 
reproducible results are obtained if the initial state is prepared using a
procedure which favors an equilibrium texture. Experimentally textural
metastability is best avoided by cooling slowly ($dT/dt \sim -1\;
\mu$K/min) through $T_{\rm c}$ at  nonzero $\Omega$. For  consistent
$v_{\rm c}$ values,  $\Omega$ must not be reduced to zero at any point during the
$v_{\rm c}$ measurements. The critical velocities measured
under these conditions are shown in Fig.\
\ref{f.selection} in axial ($\bullet$, \kern .2ex\vrule height
.95ex width .8ex depth -.15ex \kern .2ex) and in transverse field ($\circ$,
\kern .2ex\vrule height .98ex width .1ex depth -.18ex \vrule height .28ex width .7ex
depth -.18ex 
\kern -.8ex \vrule height .98ex width .7ex depth -.88ex
\vrule height .98ex width .1ex depth -.18ex\kern .2ex).  

Insets (a) and (b) in Fig.\ \ref{f.selection} display two candidates for the
$\hat{\bf l}$ texture  in a rotating cylinder in axial field. The ``circular''
texture (a) is symmetric in rotations around the cylinder axis but the
``double-half'' (b) has only reflection symmetry. In both textures the boundary
condition fixes $\hat{\bf l}$ perpendicular to the wall but the counterflow bends it
azimuthal already at distances $\sim\xi_{\rm d}$. Concerning the flow properties,
the circular texture is equivalent to the uniform texture, and thus vortices are
here expected to be created at $v_{\rm c2}$. In contrast, the double-half texture
contains two locked solitons, which means that vortices are created already at 
$v_{\rm cLS}$. The fact that the measured 
$v_{\rm c}$ data in Fig.~\ref{f.selection} lies  consistently above $v_{\rm cLS}$
suggests that the circular texture is created in cooling through $T_{\rm c}$ in
rotation. Thus the measurement
of $v_{\rm c}$ gives information about textures which is difficult to get by other
techniques \cite{Hook}.

In transverse field the texture depicted in inset (c) is expected. The flow is
here parallel to the field in two sectors of the cylinder.
There it becomes unstable toward vortex formation at  $v_{\rm
c\parallel}=v_{\rm d}\rho_\parallel/
\sqrt{\rho_\perp(\rho_\perp-\rho_\parallel)}$, as calculated by Fetter
\cite{F81}.  The measured data 
($\circ$,
\kern .2ex\vrule height .98ex width .1ex depth -.18ex \vrule height .28ex width .7ex
depth -.18ex 
\kern -.8ex \vrule height .98ex width .7ex depth -.88ex
\vrule height .98ex width .1ex depth -.18ex\kern .2ex,
{\small$\triangle$}) agrees
with this prediction.  

Fig.\ \ref{f.selection} also displays $v_{\rm c}$ measurements ($\times$) in the 
presence of an unlocked soliton, whose plane is perpendicular to
${\bf\Omega}$. It is identified from a characteristic satellite peak in the
NMR spectrum \cite{VSexp}. Also this object seems to reduce $v_{\rm c}$ to a low
but well defined value.

The measurements of $v_{\rm c}$ can also be carried out as a function of the
magnetic field. The procedure is to accelerate in the desired field, after which
the field is changed to the NMR value at constant $\Omega$, and then $v_{\rm c}$
is determined by decelerating to the annihilation threshold ($\Omega_{\rm min}$
in Fig.\ \ref{fig:AccDec}). The method works if $v_{\rm c}(H)$ does not increase
with decreasing $H$.  Indeed we find (axial field, 33.9 bar, $T/T_{\rm c} =
0.75$ -- 0.85) that $v_{\rm c}$ starts to decrease below about 3 mT, but remains
finite down to zero field: $v_{\rm c}(H=0) = 0.2$ --  0.6 mm/s. This is somewhat
lower than deduced by Saundry {\it et al.}\ from their measured critical flow
rate at temperatures $T/T_{\rm c} > 0.9$ \cite{Saundry}. 
 
{\it Calculations.}---The critical velocities were calculated assuming that the
order parameter depends only on one coordinate $x$. The energy functional
(\ref{e.f}) was minimized numerically for different values of the drive velocity
$v_{\rm n}$. The values of the parameters are from  Ref.~\cite{VSexp}, except
that we use $g_{\rm d}$ determined from the NMR shift in the B phase,
corrected by trivial strong-coupling effects. No adjustable parameters are
contained in the calculations. 

For helical textures the minimization gives the energy $F(v_{\rm n},q)$, where
$q$ is the fundamental wave vector of the helix. Although the periodic boundary
conditions in the rotating container tend to fix $q$, our simulation of the
dynamics gives transitions from one value of $q$ to another. As a consequence,
$q$ approximates $q_0(v_{\rm n})$, which corresponds to the minimum of
$F(v_{\rm n},q)$. (For stability both the eigenvalues of the second derivative
matrix of $F(v_{\rm n},q)$ have to be positive.) At $v_{\rm c2}$ the opening
angle $\beta(x)$ locally starts to grow larger than
$\beta\lesssim 60^\circ$ found for stable helixes and eventually sweeps through
$180^\circ$.  Previously $v_{\rm c1}$ and an estimate of $v_{\rm c2}$ have been
calculated near $T_{\rm c}$ by Lin-Liu {\it et al.} \cite{LVM}. We find that in
high field (Fig.\ 2) the helical texture is not stable above
$0.8T_{\rm c}$, but at lower temperatures $v_{\rm c2}$ grows considerably above
$v_{\rm c1}$.

In low fields ($H\lesssim H_{\rm d}$) $v_{\rm c1}$ drops 
and vanishes at $H=0$ when $T<0.85 \, T_{\rm c}$, while $v_{\rm c2}$ is
nearly independent of $H$. The observed reduction of the measured $v_{\rm
c}$ probably arises because the texture at small $H$ becomes more susceptible to
different perturbations such as heat flows or end plates of the cylinder.  

In the presence of solitons, the precession  of the whole soliton texture around
$x$ leads to phase slippage. In zero field, both $v_{\rm cLS}$ and $v_{\rm cUS}$
vanish because there is nothing to resist the precession. Such precessing states
of both the LS and the US have been studied in Refs.~\cite{HH,DH}. In contrast,
a field ${\bf H}$ perpendicular to $x$ gives rise to a finite $v_{\rm c}$, as
calculated approximately for the US by Vollhardt and Maki \cite {VM,VolWol}. We
find a much lower $v_{\rm cUS}$ than reported previously. 

{\it Conclusions.}---Our measurements of vortex formation in
$^3$He-A are the first to allow detailed comparison with theoretical
calculations.  The quantitative agreement is much
better than found for $v_{\rm c}$ in other superfluids ($^4$He-II and $^3$He-B).
We find that the critical velocity depends on the bulk $\hat{\bf l}({\bf r})$
texture. The maximal critical velocity we associate with the equilibrium
texture in axial field, while the minimal velocity is a characteristic of
textures incorporating unlocked solitons. 

We thank R. H\"anninen and J. Ruohio for valuable help. This work
is funded by the EU Human Capital and Mobility Program (no. CHGECT94-0069).


\begin{references}
\bibitem{CUV} The vortices under these conditions are of ``continuous
unlocked'' type with $\kappa=2(h/ 2m)$ \cite{AphDiag}.
\bibitem{SplaySol} Strictly taken, $v_{\rm cUS}$ is the theoretical critical
velocity for the initial growth of a vortex sheet (measured in Ref.\
\protect\cite{VSexp}) rather than for the nucleation of vortex lines.
\end{references}
\end{document}